\documentclass[twocolumn,aps]{revtex4}
\usepackage{epsfig}
\usepackage{amsmath}

\newcommand{\be}{\begin{equation}}
\newcommand{\ee}{\end{equation}}
\newcommand{\bea}{\begin{eqnarray}}
\newcommand{\eea}{\end{eqnarray}}

\begin{document}
\title{Electroweak Model Without A Higgs Particle}
\author{J. W. Moffat}
\address{Perimeter Institute for Theoretical Physics, 31
Caroline St North, Waterloo N2L 2Y5, Canada\\
and\\  Department of Physics, University of Waterloo, Waterloo,
Ontario, Canada\\}

\begin{abstract}
An electroweak model is formulated in a finite, four-dimensional
quantum field theory without a Higgs particle. The W and Z masses
are induced from the electroweak symmetry breaking of one-loop
vacuum polarization graphs. The theory contains only the observed
particle spectrum of the standard model. In terms of the observed
twelve lepton and quark masses, a loop calculation of the
non-local electroweak energy scale $\Lambda_W$ and $\rho$ predicts
the values $\Lambda_W(M_Z)=541$ GeV and $\rho(M_Z)=0.993$,
yielding $s^2_Z\equiv\sin^2\theta_W(M_Z)=0.21686\pm 0.00097$.
Possible ways to detect a non-local signal in scattering
amplitudes involving loop graphs at the LHC are discussed. Fermion
masses are generated from a ``mass gap'' equation obtained from
the lowest order, finite fermion self-energy with a broken
symmetry vacuum state. The cross section for $W_L^+
W_L^-\rightarrow W_L^+ W_L^-$ is predicted to vanish for $\sqrt{s}
> {1\,\rm TeV}$, avoiding a violation of the unitarity bound. The
Brookhaven National Laboratory measurement of the anomalous
magnetic moment of the muon and the residual difference between
the measured value and the standard model can provide a test of a
non-local deviation from the standard model. \vskip 0.25 true in
e-mail: john.moffat@utoronto.ca \vskip 0.25 true in
\end{abstract}

\pacs{PACS Numbers: *** } 
\date{\today}

\maketitle

\section{Introduction}

Although the standard model has been remarkably successful, as of
2007 no experiment has directly detected the existence of the
Higgs boson. After more than forty years of particle physics, we
are faced with the fundamental question: {\it What breaks
electroweak symmetry?} The answer is expected to be provided by
the Large Hadron Collider (LHC) at CERN, which will begin
operations in 2008.

The Higgs field in vacuum acquires a non-zero value with a
constant vacuum expectation value equal to $246$ GeV, which
spontaneously breaks the electroweak gauge symmetry $SU(2)\times
U(1)$~\cite{Higgs}. The Higgs mechanism gives mass to the gauge bosons and to
the observed leptons and quarks in the standard
model~\cite{Halzen,Burgess}. The non-observation of clear signals
leads to an experimental lower bound for the Higgs mass of $114.4$
GeV at 95\% confidence level. The standard model does not predict
the Higgs mass but if the mass is between 115 and 180 GeV, then
the standard model can remain valid up to the Planck energy scale
$\sim 10^{16}$ TeV. It is expected from theoretical arguments that
the highest possible mass allowed for the Higgs boson is around
$\sim\, 0.8 - 1$ TeV. Supersymmetry models predict that the
lightest Higgs boson of several such bosons should have a mass
around 120 GeV. The LEP Working Group predicts the Higgs mass to
be around $m_H\sim 144$ GeV based on precision electroweak data,
non-observation of the Higgs today and the hypothesis that the
minimal standard model is correct~\cite{CERN}. It is expected that
the LHC will be able to confirm or deny the existence of the Higgs
particle.

In the following, an electroweak model based on a finite gauge
field theory~\cite{Moffat,Moffat2,Moffat3,Moffat4,Moffat5} begins
with an initially $SU(2)_L\times U(1)_Y$ gauge invariant and
massless theory for free non-interacting particles. Then a finite,
non-local interaction possessing an extended gauge symmetry with
the infinite-dimensional gauge group $G(NL,4)$ makes the theory
finite to all orders of perturbation
theory~\cite{Moffat5,Hand,Woodard,Kleppe,Hand2,Cornish,Cornish2,Cornish3,Woodard2,Clayton,
Paris,Troost,Joglekar,Moffat6,Moffat7}. A dynamical electroweak
symmetry breaking measure in the path integral breaks the
$G(NL,4)$ gauge symmetry. The symmetry breaking mechanism allows
the $W$ and $Z$ gauge bosons to acquire masses through the lowest
order vacuum polarization graphs containing fermion loops, while
the photon remains massless. The tree graphs are identical to the
standard model, excluding the Higgs particle and they are strictly
local, maintaining classical locality and macroscopic causality.
The non-local effects in the theory occur only in quantum loops
and vanish as $\hbar\rightarrow 0$. An experimental signature for
the model is that the cross section for $W^+_LW^-_L\rightarrow
W^+_LW^-_L$ ($W_L$ denotes the longitudinal part of the massive
intermediate vector boson $W$) vanishes above $\sim\, 1$ TeV
avoiding a violation of the unitarity bound. Thus, the W and Z
bosons become massless as the measure becomes the $G(NL,4)$ gauge
invariant measure above $\sim\, 1$ TeV, and only the transverse
components of the W and Z bosons survive.

An important signature for discovering the origin of electroweak
symmetry breaking is the observation at the LHC of $WW$
scattering. The vanishing of the $W_LW_L\rightarrow W_LW_L$
scattering cross section above $\sim\,1$ TeV would be a signature
for a no-Higgs particle, ultra-violet finite quantum field theory.
On the other hand, if the cross section is observed to be strong
above $\sim\, 1$ TeV, then it is saying that possible new strong
interactions and the exchange of new intermediary particles are
responsible for the electroweak symmetry breaking. If the Higgs
particle is observed at the LHC, then the $W_LW_L$ scattering will
not be strong, but at the same time it will not be expected to
vanish above $\sim\, 1$ TeV.

The masses of fermions are generated through the mass-gap equation
obtained from the lowest order finite, fermion-boson self-energy
graph with a broken symmetry vacuum state. The spectrum of
particles only contains the observed $W$, $Z$ and photon particles
and the standard quarks and leptons.

The electroweak model based on a finite quantum field theory
(FQFT) without a Higgs particle avoids the fine-tuning (hierarchy)
problem associated with the Higgs scalar field radiative
corrections. A calculation of the $\rho$ parameter from the finite
boson-fermion self-energy loop graphs yields the predictions for
the weak mixing angle, $\sin^2\theta_W(M_Z)=0.21686\pm 0.00097$
and the electroweak non-local energy scale
$\Lambda_W(M_Z)=541$ GeV. A calculation of the muon $g-2$
anomalous magnetic moment in the FQFT can provide a means for
detecting a non-local deviation from the standard model in
perturbative loop diagrams.

\section{Finite Non-Local Electroweak Theory}

We shall choose units $c=\hbar=1$ and the metric ${\rm
diag}(\eta_{\mu\nu})=(-1,+1,+1,+1)$. The Lagrangian takes the
form:
\begin{equation}
\label{Lagrangian} L=L_0+L_B+{\cal L}_I+\tilde{\cal L}_I,
\end{equation}
where $L_0$ is the local, free kinetic Lagrangian for massless
leptons and quarks given by
\begin{equation}
L_0=[{\bar\psi}^L(x)i\gamma\cdot\partial\psi^L(x)
+{\bar\psi}^R(x)i\gamma\cdot\partial\psi^R(x)],
\end{equation}
where $\gamma\cdot\partial=\gamma^\mu\partial_\mu$. The fields
$\psi^L(x)$ and $\psi^R(x)$ denote {\it local} two-component
left-handed lepton and quark doublet fields and right-handed
lepton and quark singlet fields, respectively, with
$\psi^L=(1/2)(1-\gamma_5)\psi$ and $\psi^R=(1/2)(1+\gamma_5)\psi$.
The local boson Lagrangian density $L_B$ is given by
\begin{equation}
L_B=-\frac{1}{4}G^{a\mu\nu}
G_{a\mu\nu}-\frac{1}{4}B^{\mu\nu}B_{\mu\nu},
\end{equation}
where
\begin{equation}
G^{\mu\nu}_a=\partial^\mu W^\nu_a-\partial^\nu
W^\mu_a+g\epsilon_{abc}W_b^\mu W_c^\nu,
\end{equation}
and
\begin{equation}
B_{\mu\nu}=\partial_\mu B_\nu-\partial_\nu B_\mu.
\end{equation}

The $\tilde{\cal L}_I$ is an iteratively defined series of higher
interactions which strip the non-locality from the tree graphs.
The {\it non-local} interaction Lagrangian density is described by
\begin{equation}
{\cal L}_I=-g{\cal J}^\mu_a(x){\cal W}_{a\mu}(x)-g'{\cal
J}^\mu_Y(x){\cal B}_\mu(x),
\end{equation}
where $g$ and $g'$ are electroweak coupling constants. The ${\cal
W}_{a\mu}$ and ${\cal B}_\mu$ are the non-local gauge boson
fields, while the ${\cal J}^\mu_a$ and ${\cal J}_Y$ are the
non-local weak isospin and hypercharge currents:
\begin{equation}
{\cal J}^\mu_a(x)
=\frac{1}{2}{\bar\Psi}^L(x)\gamma^\mu\tau_a\Psi^L(x)\quad (a=1,2,3),
\end{equation}
and
\begin{equation}
{\cal J}^\mu_Y(x) =-\frac{Y}{2}{\bar\Psi}^L(x)\gamma^\mu\Psi^L(x)-
\frac{Y}{2}{\bar\Psi}^R(x)\gamma^\mu\Psi^R(x),
\end{equation}
where $\tau_a$ denote the Hermitian Pauli matrices and $Y$ denotes
the hypercharge. The $\Psi$ denote the non-local lepton and quark
fields in the ${\cal J}^\mu_a$ and ${\cal J}^\mu_Y$ currents. In
the absence of interactions, ${\cal L}_I=0$, the massless
Lagrangian is invariant under $SU(2)\times U(1)$ gauge
transformations.

The $A_\mu$ and $Z_\mu$ are linear combinations of the two fields
$W_{3\mu}$ and $B_\mu$:
\begin{equation}
\label{A} A_\mu=\cos\theta_WB_\mu+\sin\theta_WW_{3\mu},
\end{equation}
\begin{equation}
\label{Z} Z_\mu=-\sin\theta_WB_\mu+\cos\theta_WW_{3\mu},
\end{equation}
where the angle $\theta_W$ denotes the weak mixing angle. The
electroweak coupling constants $g$ and $g'$ are related to the
electric charge $e$ by the standard equation
\begin{equation}
g\sin\theta_W=g'\cos\theta_W=e
\end{equation}
 and we use the standard normalization $\cos\theta_W=g/(g^2+g^{'2})^{1/2}$ and
 $g'/g=\tan\theta_W$.

The non-local field operators ${\cal W}_{a\mu},{\cal B}_\mu$ and
$\Psi$ are defined in terms of the local operators
$W_{a\mu},B_\mu$ and $\psi$ by
\begin{equation}
{\cal W}_{a\mu}=\int d^4y
G(x-y)W_{a\mu}(y)=G\biggl(\frac{\partial^2}{\Lambda^2}\biggr)W_{a\mu}(x),
\end{equation}
\begin{equation}
{\cal B}_\mu(x)=\int d^4y
G(x-y)B_{\mu}(y)=G\biggl(\frac{\partial^2}{\Lambda^2}\biggr)B_{\mu}(x),
\end{equation}
and
\begin{equation}
\Psi(x)=\int d^4y G(x-y)\psi(y)=G\biggl(\frac{\partial^2}{\Lambda^2}\biggr)\psi(x).
\end{equation}
Here, $\partial^2=\partial^\mu\partial_\mu$ and $G(\partial^2/\Lambda^2)$ is
a Lorentz-invariant operator distribution whose momentum space
Fourier transform {\it is an entire function}. We can write
\begin{equation}
G(x-y)=G\biggl(\frac{\partial^2}{\Lambda^2}\biggr)\delta^4(x-y),
\end{equation}
where $\ell\sim1/\Lambda$ is a small invariant interval and the
high Euclidean momentum damping means that the non-locality has
compact support. The Fourier transform of the damping function is
defined by
\begin{equation}
\int
d^4x\exp(ipx)G\biggl(\frac{\partial^2}{\Lambda^2}\biggr)
=K\biggl(-\frac{p^2}{\Lambda^2}\biggr).
\end{equation}
Unitarity conditions require that $K(-p^2/\Lambda^2)$ is an entire
function that does not generate new poles on the physical sheet
and that the residue of the physical pole remains
unity~\cite{Moffat5}.

The non-local distribution operator is defined by
\begin{equation}
G\biggl(\frac{\partial^2}{\Lambda_W^2}\biggr)\equiv {\cal
E}_m=\exp\biggl(\frac{\partial^2-m^2}{2\Lambda_W^2}\biggr),
\end{equation}
where $\Lambda_W$ denotes the electroweak non-local energy scale.
We have
\begin{equation}
\Psi={\cal E}_0\psi,\quad \bar\Psi={\cal E}_0\bar\psi,
\end{equation}
\begin{equation}
{\cal W}_{a\mu}={\cal E}_0W_{a\mu},\quad {\cal B}_\mu={\cal
E}_0B_\mu.
\end{equation}

Consider, as an example, the stripping of the non-locality of the
tree graphs and the restoration of gauge invariance for the
$B_\mu$ coupling. We have~\cite{Moffat5}:
\begin{equation}
\label{0+1} {\cal L}_{0+1}
=-\frac{1}{4}B^{\mu\nu}B_{\mu\nu}-\bar\psi
i\gamma\cdot\partial\psi +g'{\bar\Psi}\gamma\cdot {\cal B}\Psi,
\end{equation}
where we have for simplicity ignored the left and right-handed
structure of the non-local hypercharge current ${\cal J}^\mu_Y$.
Eq.(\ref{0+1}) is invariant up to order $g^{'2}$ under the
transformation:
\begin{equation}
\delta B_\mu\equiv \delta_0 B_\mu=-\partial_\mu\theta,
\end{equation}
\begin{equation}
\delta_1\psi=ig'{\cal E}_0{\hat\theta\Psi},
\end{equation}
where $\hat\theta={\cal E}_0\theta$. Here, the operator ${\cal
E}_0$ acts on the product ${\hat\theta\Psi}$, while the ${\cal
E}_0$ in $\hat\theta$ acts only upon $\theta$ and the ${\cal E}_0$
in $\Psi$ acts only upon $\psi$.

We now form the operator ${\cal O}$:
\begin{equation}
{\cal O}\equiv\frac{({\cal E}_0)^2-1}{\partial^2}
=\int_0^1\frac{d\tau}{\Lambda_W^2}\exp\biggl[\tau\frac{\partial^2}{\Lambda_W^2}\biggr].
\end{equation}
The operator ${\cal O}$ is an entire function of $\partial^2$, so it
does not produce any poles in the momentum representation that will
violate unitarity.

By using the operator ${\cal O}$, we can express the simplest
non-local four-point interaction as
\begin{equation}
{\cal L}_2=-g^{'2}\bar\Psi\gamma\cdot{\cal
B}i\gamma\cdot\partial{\cal O}\gamma\cdot {\cal B}\Psi.
\end{equation}
The scattering amplitude computed from ${\cal L}_{0+1+2}$ is
unchanged from its local point particle antecedent. This comes
about because each $V_2$ vertex contribution to the amplitude can
be split into two terms through decomposing the operator ${\cal
O}$ into ${\cal E}_0^2/\partial^2$ and $-1/\partial^2$. The first
such term cancels the contribution from the corresponding
$V_1\cdot V_1$ channel, while the second term is the local, point
particle contribution for that channel. This process can be
extended to higher $B_\mu$ amplitudes with interactions of the
form:
\begin{equation}
{\cal L}_n =-(-g')^n\bar\Psi\gamma\cdot {\cal
B}[i\gamma\cdot\partial{\cal O}\gamma\cdot{\cal B}]^{(n-1)}\Psi.
\end{equation}
This sums to give the total $B^\mu$ coupling Lagrangian:
\begin{eqnarray}
&&{\cal L} =-\frac{1}{4}B^{\mu\nu}B_{\mu\nu}-\bar\psi
i\gamma\cdot\partial\psi\nonumber\\
&& +g'\bar\Psi\gamma\cdot {\cal B}[1+g'i\gamma\cdot\partial{\cal
O}\gamma\cdot{\cal B}]^{-1}\Psi.
\end{eqnarray}
The extended tree graph scattering amplitude is the same as the
local, point particle amplitude and the decoupling of unphysical
modes is accomplished. The true amplitudes that differ from the
point particle ones contain an internal $B_\mu$ line, which are
enhanced by an exponential damping factor for each internal
momentum.

A modification of the fermionic transformation at each order is
\begin{equation}
\delta_n\psi=-i(-g')^n{\cal
E}_0\hat\theta[i\gamma\cdot\partial{\cal O}\gamma\cdot{\cal
B}]^{n-1}\Psi.
\end{equation}
Moreover, the sum of all variations gives
\begin{equation}
\delta B_\mu=-\partial_\mu\theta,
\end{equation}
\begin{equation}
\delta\psi=ig'{\cal E}_0\hat\theta[1+g'i\gamma\cdot\partial{\cal
O}\gamma\cdot{\cal B}]^{-1}\Psi.
\end{equation}
It can be proved that $\delta L=0$ at order $g^{'n}$ establishing
the gauge invariance under the non-local gauge
transformations~\cite{Moffat5}.

The non-localization of the interaction Lagrangian has resulted in
gauge transformations that mix gauge indices and spinor indices at
different spacetime coordinates. The action for the $B_\mu$
coupling is invariant under a transformation of the form
\begin{equation}
\delta B_\mu(x)=-\partial_\mu\theta(x),
\end{equation}
\begin{equation}
\delta\psi(x)=ig'\int d^4yd^4z{\cal T}[g'B](x,y,z)\theta(y)\psi(z).
\end{equation}
Here, ${\cal T}\sim 1+g'B+...$ is a representation operator that
is a spinorial matrix as well as a functional of the vector field
$B_\mu$:
\begin{equation}
{\cal T}[g'B](x,y,z)={\cal
E}_0[\delta^4(x-y)][1+g'i\gamma\cdot\partial{\cal O}\gamma\cdot{\cal
B}]^{-1}{\cal E}_0\delta^4(x-z).
\end{equation}

The transformations do not form a group, because although the
gauge group for the $B_\mu$ field is Abelian on shell, it does not
close on commutation unless the fermion fields obey their
equations of motion:
\begin{eqnarray}
&&[\delta_{\theta_1},\delta_{\theta_2}]\psi=-g^{'2}{\cal
E}_0\{{\hat\theta}_1[1+g'{\cal B}{\cal
O}i\gamma\cdot\partial]^{-1}\nonumber\\
&& \times i\gamma\cdot\partial{\cal
O}{\hat\theta}_2-(1\longleftrightarrow 2)\}{\cal
E}_0(i\gamma\cdot\partial+g'{\cal E}_0\gamma\cdot{\cal B}\nonumber\\
&& \times[1+g'\gamma\cdot{\cal B}{\cal
O}i\gamma\cdot\partial]^{-1}{\cal E}_0)\psi.
\end{eqnarray}
However, the transformations are part of an infinite-dimensional
group $G(NL,4)$ which includes transformations that vanish in the
local limit and only influence the fermi fields.

The non-localization process guarantees gauge invariance to all
orders and removes all unphysical couplings to longitudinal vector
bosons. It can be extended to the total $W$, $B$ and $\psi$
electroweak Lagrangian. The fact that the tree graphs of the
theory are the {\it purely local point-like graphs} protects the
classical theory from any violation of macroscopic causality. The
non-locality resides only in the loop sectors where a violation of
micro-causality is potentially hidden by the uncertainty
principle. The loop graphs in FQFT are finite to all orders in
perturbation theory.

We could equally well have formulated our non-local Lagrangian as
\begin{equation}
{\cal L}={\cal L}_0+{\cal L}_B+L_I+{\tilde L}_I,
\end{equation}
where now the non-locality is in the free and kinetic energy parts
of the Lagrangian and not in the interaction part $L_I$. This will
place the non-local form factor on the propagators, whereas in the
previous process the non-local form factors were imposed on the
vertices. Both processes produce a finite, gauge invariant and
unitary QFT to all orders in perturbation theory. Another method
is to introduce shadow fields and shadow
propagators~\cite{Woodard,Kleppe,Woodard2}. This method has been
successfully applied to Yang-Mills Lagrangians and to quantum
gravity~\cite{Cornish,Cornish2,Cornish3,Woodard2,Moffat6}.

We have demonstrated a systematic way of maintaining the gauge
invariance of the classically, initially non-local massless
Lagrangian which consists of two stages. In the first stage, an
$SU(2)\times U(1)$ gauge invariant interaction-free action is made
non-local and then an infinite series of chosen higher
interactions $\tilde{\cal L}_I$ is added to the Lagrangian. These
added interactions provide the theory with a new nonlinear and
non-local gauge invariance which makes Lorentz invariance
compatible with perturbative unitarity at tree order. The second
stage consists of finding a measure which makes the functional
path integral formalism invariant under the non-local gauge
symmetry without destroying perturbative unitarity, namely, by
finding a measure whose interactions are {\it entire} functions of
the derivative operator. This measure then yields a functional
formalism which defines a set of Green functions which are
ultraviolet finite and Poincar\'e invariant to all orders, and
gives scattering amplitudes which are perturbatively finite. The
scattering amplitudes are then analytically continued into the
Euclidean momentum space.

\section{Path Integral Formalism and Measure Factors}

The path integral formalism is completed with the expression:
\begin{equation}
\langle T^*(O)\rangle =\int[D\bar\psi][D\psi][DW][DB]\mu_{\rm
inv}O\exp(iS),
\end{equation}
where $O$ is a given operator. All the loop graphs are ultraviolet
finite and unitary to all orders of perturbation theory for the
non-local gauge invariant Lagrangian. In the limit that the
non-local weak scale $\Lambda_W\rightarrow \infty$, the path
integral formalism becomes that of the renormalizable, local point
field theory of massless gauge bosons $W$ and $B$.

A determination of measures for fermion loops including all
lepton, quark, parity and isospin contributions to vacuum
polarization loops has been obtained~\cite{Moffat3}. For the
$W^\pm$ sector, the invariant measure for the fermion loops is
given by
\begin{equation}
\ln\biggl(\mu_{\rm inv}[W^\pm]\biggr)=ig^2\int d^4x{\cal
W}^{+\mu}{\cal M}_{\mu\nu}[W^\pm]{\cal W}^{-\mu},
\end{equation}
where
\begin{eqnarray}
&&{\cal
M}_{\mu\nu}[W^\pm]=\frac{\eta_{\mu\nu}}{(4\pi)^2}\sum_dC_d(L_{1+}(m_1,m_2)\nonumber\\
&& +L_{2+}(m_1,m_2) +L_{2+}(m_2,m_1)).
\end{eqnarray}
Here the sum is over all fermion doublets $d$, and $C_d$ denotes
the color factors.

The $W^{3\mu}$ and $B^\mu$ gauge boson invariant measures for the
fermion loop sector are given by
\begin{equation}
\ln\biggl(\mu_{\rm inv}[W^3]\biggr)=\frac{ig^2}{2}\int d^4x{\cal
W}^{3\mu}{\cal M}_{\mu\nu}[W^3]{\cal W}^{3\nu},
\end{equation}
\begin{equation}
{\cal M}_{\mu\nu}[W^3]
=\frac{\eta_{\mu\nu}}{2(4\pi)^2}\sum_\mathrm{lhf}C_f(L_{1+}(m,m)+2L_{2+}(m,m)),
\end{equation}
and
\begin{equation}
\ln\biggl(\mu_{\rm inv}[B]\biggr)=\frac{ig^{'2}}{2}\int d^4x{\cal
B}^\mu{\cal M}_{\mu\nu}[B]{\cal B}^\nu,
\end{equation}
\begin{eqnarray}
&&{\cal M}_{\mu\nu}[B]
=2\frac{\eta_{\mu\nu}}{(4\pi)^2}\biggl(\sum_fC_f
\frac{Y_f^2}{4}(L_{1+}(m,m)\nonumber\\
&&+2L_{2+}(m,m))+\sum_\mathrm{rhf}C_f\frac{Y_LY_R}{4}(2M_1(m)+M_2(m))\biggr),\nonumber\\
\end{eqnarray}
where $\sum_\mathrm{lhf}$ and $\sum_\mathrm{rhf}$ denote sums over
left-handed and right-handed fermions only, respectively.
Moreover, $\sum_f=\sum_\mathrm{rhf}+\sum_\mathrm{lhf}$, $Y_f$ denotes the
fermion hypercharge factor and $Y_L$ and $Y_R$ denote the
left-handed and right-handed hypercharge factors, respectively.
The invariant measure for the off-diagonal $W^3-B$ for the fermion
loops is
\begin{equation}
\ln\biggl(\mu_{\rm inv}[W^3-B]\biggr)=igg'\int d^4x{\cal
W}^{3\mu}{\cal M}_{\mu\nu}[W^3-B]{\cal B}^\nu,
\end{equation}
\begin{eqnarray}
&&{\cal M}_{\mu\nu}[W^3-B]
=-\frac{\eta_{\mu\nu}}{(4\pi)^2}\biggl(\frac{1}{2}\sum_\mathrm{lhf}C_f(L_{1+}(m,m)\nonumber\\
&& +2L_{2+}(m,m)) +\sum_\mathrm{rhf}C_f\frac{Y_f}{2}
(M_1(m)+M_2(m))\biggr).\nonumber\\
\end{eqnarray}

The $Ls$ are given by
\begin{eqnarray}
&&L_{1\pm}(m_1,m_2)=\int^\infty_1 d\tau_1\int^\infty_1
d\tau_2\exp\biggl[-\tau_1\frac{m_1^2}{\Lambda_W^2}-\tau_2\frac{m_2^2}{\Lambda_W^2}\nonumber\\
&&-\frac{\tau_1\tau_2}{(\tau_1+\tau_2)}\frac{p^2}{\Lambda_W^2}\biggr]
\biggl(\frac{\Lambda^2_W}{(\tau_1+\tau_2)^3}\pm\frac{p^2\tau_1\tau_2}{(\tau_1+\tau_2)^4}\biggr),
\end{eqnarray}
\begin{eqnarray}
&&L_{2\pm}(m_1,m_2)=\int_0^1 d\tau_1\int^\infty_1
d\tau_2\exp\biggl[-\tau_1\frac{m_1^2}{\Lambda_W^2}-\tau_2\frac{m_2^2}{\Lambda_W^2}\nonumber\\
&&-\frac{\tau_1\tau_2}{(\tau_1+\tau_2)}\frac{p^2}{\Lambda_W^2}\biggr]
\biggl(\frac{\Lambda^2_W}{(\tau_1+\tau_2)^3}\pm\frac{p^2\tau_1\tau_2}{(\tau_1+\tau_2)^4}\biggr).
\end{eqnarray}
We also have
\begin{eqnarray}
&&M_1=\int^\infty_1 d\tau_1\int^\infty_1
d\tau_2\frac{m^2}{(\tau_1+\tau_2)^2}\nonumber\\
&& \times\exp\biggl[-(\tau_1+\tau_2)\frac{m^2}{\Lambda_W^2}
-\frac{\tau_1\tau_2}{(\tau_1+\tau_2)}\frac{p^2}{\Lambda_W^2}\biggr],
\end{eqnarray}
\begin{eqnarray}
\lefteqn{M_2=\int^1_0 d\tau_1\int^\infty_1
d\tau_2\frac{m^2}{(\tau_1+\tau_2)^2}}\nonumber\\
&& \times\exp\biggl[-(\tau1+\tau_2)\frac{m^2}{\Lambda_W^2}
-\frac{\tau_1\tau_2}{(\tau_1+\tau_2)}\frac{p^2}{\Lambda_W^2}\biggr].
\end{eqnarray}

The transverse, gauge invariant vacuum polarization tensor for the
$W$ boson has been determined~\cite{Moffat2,Moffat3}. For the
fermion loop sector, we have
\begin{equation}
\Pi_W^{\mu\nu}(p^2)=\Pi_W^T(p^2)\biggl(\eta^{\mu\nu}-\frac{p^\mu
p^\nu}{p^2}\biggr)+\Pi_W^L(p^2)\frac{p^\mu p^\nu}{p^2},
\end{equation}
where $\Pi_W^T$ and $\Pi_W^L$ denote the transverse and
longitudinal parts, respectively. We obtain $\Pi_W$ by adding
together the three contributions $\Pi_{W1}$, $\Pi_{W2}$ and
$\Pi_{W3}$, where the first term is produced by the standard
boson-fermion loop graph, the second by the gauge boson-fermion
tadpole graph and the third by the graph associated with the
fermion measure factor. The three one-loop vacuum polarization
graphs are shown in Figure 1 and Figure 2, and the measure factor
graphs are shown in Figure 3.

\begin{figure}
\includegraphics[width=1.0\linewidth]{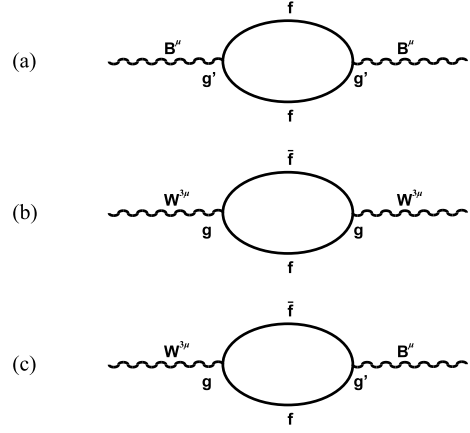}
\caption{(a) vacuum polarization fermion one-loop graph for the B
boson; (b) vacuum polarization fermion one-loop graph for the W boson; (c)
the off-diagonal B-W boson fermion one-loop vacuum polarization graph.}
\label{fig:1}
\end{figure}

\begin{figure}
\includegraphics[width=1.0\linewidth]{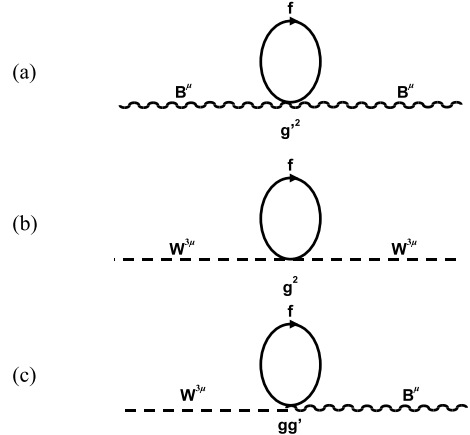}
\caption{(a) the tadpole fermion one-loop vacuum polarization
graph for the B boson; (b) the tadpole fermion one-loop vacuum polarization graph for the W boson; (c) the off-diagonal
tadpole fermion W-B boson vacuum polarization graph.}
\label{fig:2}
\end{figure}

\begin{figure}
\includegraphics[width=1.0\linewidth]{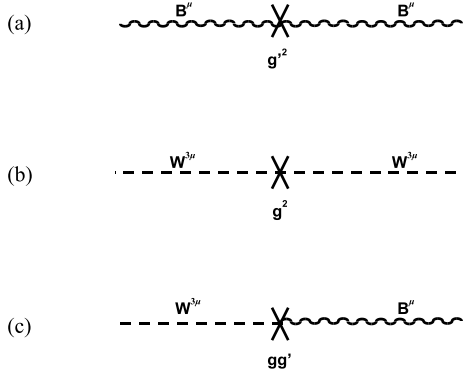}
\caption{(a) the B boson fermion measure factor contribution; (b)
the W boson fermion measure factor contribution; (c) the
off-diagonal B-W fermion measure factor contribution.}
\label{fig:3}
\end{figure}

It can be shown that for the $W^\pm$ sector,
$\Pi_W^L=\Pi^L_{W1}+\Pi^L_{W2}+\Pi^L_{W3}=0$ and the transverse
part is given by
\begin{eqnarray}
\lefteqn{\Pi_W^T(p^2)=
-\frac{g^2}{(4\pi)^2}\exp\biggl(-\frac{p^2}{\Lambda_W^2}\biggr)
\sum_dC_d}\nonumber\\
&& \times\biggl(P_1(m_1,m_2)+P_2(m_1,m_2)+P_2(m_2,m_1)\biggr).
\end{eqnarray}
The $Ps$ are given by
\begin{eqnarray}
\lefteqn{P_1=L_{1+}-L_{1-} =2p^2\int^\infty_1 d\tau_1\int^\infty_1
d\tau_2}\nonumber\\
&& \times
\exp\biggl[-\tau_1\frac{m_1^2}{\Lambda_W^2}-\tau_2\frac{m_2}{\Lambda_W^2}
-\frac{\tau_1\tau_2}{\tau_1+\tau_2}\frac{p^2}{\Lambda_W^2}\biggr]
\frac{\tau_1\tau_2}{(\tau_1+\tau_2)^4},\nonumber\\
\end{eqnarray}
and
\begin{eqnarray}
\lefteqn{P_2=L_{2+}-L_{2-} =2p^2\int^1_0 d\tau_1\int^\infty_1
d\tau_2}\nonumber\\
&& \times
\exp\biggl[-\tau_1\frac{m_1^2}{\Lambda_W^2}-\tau_2\frac{m_2}{\Lambda_W^2}
-\frac{\tau_1\tau_2}{\tau_1+\tau_2}\frac{p^2}{\Lambda_W^2}\biggr]
\frac{\tau_1\tau_2}{(\tau_1+\tau_2)^4}.\nonumber\\
\end{eqnarray}
We observe that $P_1$ and $P_2$ are proportional to $p^2$, so they
vanish as $p^2\rightarrow 0$ guaranteeing that $\Pi^T(0)=0$ and
keeping the gauge bosons $W$ and $Z$ massless.

The measures for the invariant pure Yang-Mills boson loops can be
determined. For example, the $B^\mu$ measure is given by
\begin{equation}
\ln\biggl(\mu_{\rm inv}[B]\biggr)=-ig^{'2}\int d^4x{\cal
B}^\mu{\cal M}[B]{\cal B}_\mu,
\end{equation}
where
\begin{eqnarray}
&&{\cal M}[B]
=\frac{\Lambda_W^2}{2^4\pi^2}\int^1_0\frac{d\tau}{(\tau+1)^2}\exp\biggl(-\frac{\tau}
{(\tau+1)}\frac{p^2}{\Lambda_W^2}\biggr)\nonumber\\
&& \times\biggl(\frac{2}{\tau+1}-3+6\frac{\tau}{\tau+1}\biggr).
\end{eqnarray}

\section{Dynamical Symmetry Breaking}

We must now introduce a method to generate the physical $W$ and
$Z$ gauge boson masses. This will be achieved by dynamically
breaking the electroweak non-local gauge symmetry by an
appropriate choice of the fermion sector measure, so that
$\Pi(m_{\rm phys}^2)\not=0$ and a mass is induced at $p^2=-m_{\rm
phys}^2$, making the $W$ and $B$ particles massive vector bosons.

To lowest order, we find that the $W$ boson propagator is modified
according to
\begin{equation}
D_{W\alpha\beta}(p)\rightarrow
D_{W\alpha\beta}(p)+D_{W\alpha\mu}(p)g^2\Pi_W^{\mu\nu}(p^2)D_{W\nu\beta}(p).
\end{equation}
We now have that
\begin{equation}
\frac{-\eta_{\alpha\beta}}{p^2-i\epsilon}\rightarrow\frac{-\eta_{\alpha\beta}}{p^2-i\epsilon}
+\frac{\eta_{\alpha\mu}}{p^2-i\epsilon}g^2\Pi^{\mu\nu}_W(p^2)\frac{\eta_{\nu\beta}}{p^2-i\epsilon}.
\end{equation}
It follows that
\begin{equation}
\frac{1}{p^2-i\epsilon}\rightarrow\frac{1}{p^2-i\epsilon}
-\frac{1}{p^2-i\epsilon}g^2\Pi_W(p^2)\frac{1}{p^2-i\epsilon},
\end{equation}
where
\begin{equation}
v\equiv\Pi_W(p^2)=\frac{1}{4}{\Pi_W^\mu}_\mu(p^2).
\end{equation}
To lowest order we get
\begin{equation}
\label{Bmasses} \frac{1}{p^2-i\epsilon}\rightarrow
\frac{1}{p^2+g^2\Pi_W(p^2)-i\epsilon}+O(g^4),
\end{equation}
where in our regularized theory $\Pi_W(p^2)$ is finite. The mass
is defined on the mass shell to be
\begin{equation}
m^2_{\rm phys}=g^2\Pi(m^2_{\rm phys}).
\end{equation}

The symmetry breaking measure is defined by
\begin{eqnarray}
&&\ln\biggl(\mu_{SB}\biggr)=-igg'\int d^4x{\cal
W}^{3\mu}\biggl({\cal M}_{\mu\nu}[W^3-B]\biggr)_{SB}{\cal
B}^\nu\nonumber\\
&&+\frac{ig^{'2}}{2}\int d^4x{\cal B}^\mu\biggl({\cal
M}_{\mu\nu}[B]\biggr)_{SB}{\cal B}^\nu,
\end{eqnarray}
where
\begin{eqnarray}
&&\biggl({\cal M}_{\mu\nu}[W^3-B]\biggr)_{SB} =\nonumber\\
&& -\frac{\eta_{\mu\nu}}{(4\pi)^2}\sum_\mathrm{rhf}C_f\frac{Y_f}{2}
(M_1(m)+M_2(m)),
\end{eqnarray}
and
\begin{eqnarray}
&&\biggl({\cal M}_{\mu\nu}[B]\biggr)_{SB}
=\frac{2\eta_{\mu\nu}}{(4\pi)^2}\biggl\{\sum_\mathrm{rhf}
C_f\bigg[\frac{Y_f^2}{4}(L_{1+}(m,m)\nonumber\\
&& +2L_{2+}(m,m))
+\frac{Y_LY_R}{4}(2M_1(m)+M_2(m))\biggr]\nonumber\\
&& +\sum_\mathrm{lhf}C_f\biggl(\frac{Y_f^2-1}{4}\biggr)
(L_{1+}(m,m)+2L_{2+}(m,m))\biggr\}.\nonumber\\
\end{eqnarray}

The vacuum polarization tensor $\Pi^{\mu\nu}(W^3)$ for the $W^3$
sector is given by
\begin{eqnarray}
&&\Pi^{\mu\nu}(W^3)=\frac{g^2}{2(4\pi)^2}\exp\biggl(-\frac{p^2}{\Lambda_W^2}\biggr)
\sum_\mathrm{lhf}C_f\biggl[(L_{1+}(m,m)\nonumber\\
&& +2L_{2+}(m,m))\biggl(\eta^{\mu\nu}-\frac{p^\mu
p^\nu}{p^2}\biggr)\nonumber\\
&& +(L_{1-}(m,m)+2L_{2-}(m,m))\frac{p^\mu p^\nu}{p^2}\biggr].
\label{W3}
\end{eqnarray}
For the $W^\pm$ sector we have
\begin{eqnarray}
&&\Pi^{\mu\nu}(W^\pm)=\frac{g^2}{2(4\pi)^2}\exp\biggr(-\frac{p^2}{\Lambda_W^2}\biggr)
\sum_dC_d\biggl[(L_{1+}(m_1,m_2)\nonumber\\
&& +L_{2+}(m_1,m_2)+L_{2+}(m_2,m_1))\biggl(\eta^{\mu\nu}-\frac{p^\mu
p^\nu}{p^2}\biggr)\nonumber\\
&& +(L_{1-}(m_1,m_2)+L_{2-}(m_1,m_2)+L_{2-}(m_2,m_1))\frac{p^\mu
p^\nu}{p^2}\biggr].\nonumber\\\label{Wcharged}
\end{eqnarray}

We get for the $B$ and $W^3-B$ sectors
\begin{eqnarray}
&&\Pi^{\mu\nu}(B)=\frac{g^{'2}}{2(4\pi)^2}\exp\biggl(-\frac{p^2}{\Lambda_W^2}\biggr)
\sum_\mathrm{lhf}C_f\nonumber\\
&&\times\biggl[(L_{1+}(m,m)+2L_{2+}(m,m))\biggl(\eta^{\mu\nu}-\frac{p^\mu
p^\nu}{p^2}\biggr)\nonumber\\
&& +(L_{1-}(m,m)+2L_{2-}(m,m))\frac{p^\mu p^\nu}{p^2}\biggr],
\end{eqnarray}
and
\begin{eqnarray}
&&\Pi^{\mu\nu}(W^3-B)=-\frac{gg'}{2(4\pi)^2}\exp\biggl(-\frac{p^2}{\Lambda_W^2}\biggr)
\sum_\mathrm{lhf}C_f\nonumber\\
&&\times\biggl[(L_{1+}(m,m)+2L_{2+}(m,m))\biggl(\eta^{\mu\nu}-\frac{p^\mu
p^\nu}{p^2}\biggr)\nonumber\\
&&+(L_{1-}(m,m)+2L_{2-}(m,m))\frac{p^\mu p^\nu}{p^2}\biggr].
\end{eqnarray}

The measure $\mu_{SB}$ dynamically breaks the non-local gauge
invariance and the $W$ and $B$ gauge bosons acquire a finite mass.
The relative strength of the neutral and charged current
interactions is fixed by means of the standard relation:
\begin{equation}
\label{rhoratio}
\rho\frac{G_F}{\sqrt{2}}=\frac{g^2}{8M_Z^2\cos^2\theta_W},
\end{equation}
where $G_F$ is the Fermi constant and $G_F/\sqrt{2}=g^2/8M_W^2$.
We obtain from (\ref{rhoratio}) the standard result:
\begin{equation}
\label{rho} \rho=\frac{M_W^2}{M_Z^2\cos^2\theta_W}.
\end{equation}

The Lagrangian picks up a finite mass contribution for $\rho=1$
from the total sum of polarization graphs:
\begin{eqnarray}
\lefteqn{L_M =\frac{1}{8}v^2g^2[(W^1_\mu)^2+(W^2_\mu)^2]}\nonumber\\
&&+\frac{1}{8}v^2[g^2(W^3_\mu)^2-2gg'W^3_\mu
B^\mu+g^{'2}B^2_\mu]\nonumber\\
&&=\frac{1}{4}g^2v^2W^+_\mu W^{-\mu}\nonumber\\
&&+\frac{1}{8}v^2(W_{3\mu},B_\mu)\left(\begin{matrix}g^2 &
-gg'\\-gg' &
g^{'2}\end{matrix}\right)\left(\begin{matrix}W^{3\mu}\\B^\mu\end{matrix}\right),
\label{massmatrix}
\end{eqnarray}
where $W^{\pm}_\mu=(W^1\mp iW^2)/\sqrt{2}$  and
we can fix the value of $v$ from $G_F/\sqrt{2}=g^2/8M_W^2$ to be
$v=246$ GeV. We see that we have the usual dynamical symmetry
breaking mass matrix in which one of the eigenvalues of the
$2\times 2$ matrix in (\ref{massmatrix}) is zero. From (\ref{A})
and (\ref{Z}), we get for $\rho=1$:
\begin{equation}
\label{WZ} M_W=\frac{1}{2}vg,\quad
M_Z=\frac{1}{2}v(g^2+g^{'2})^{1/2},\quad M_A=0.
\end{equation}

\section{Evaluation of $\rho$ and $\Lambda_W$}

We shall now calculate $\rho$ and $\Lambda_W$ from the loop
diagrams for non-zero lepton and quark masses, $m_f\not=0$. We
obtain for $\rho$ the result:
\begin{eqnarray}
\lefteqn{\rho=2\sum_iC_id_i[L_1(p^2,m_{dbi1},m_{dbi2})}\nonumber\\
&&+L_2(p^2,m_{dbi1},m_{dbi2})+L_2(p^2,m_{dbi2},m_{dbi1})]\nonumber\\
&&\times\{\sum_iC_if_i[L_1(p^2,m_{fi})+2L_2(p^2,m_{fi})]\}^{-1}.
\label{rhoeq}
\end{eqnarray}
Using $M_W=(1/2)vg$, we can obtain an equation that can be solved for the electroweak energy
scale $\Lambda_W$:
\begin{eqnarray}
\lefteqn{v= \frac{1}{4\pi} \{\exp(-p^2/\Lambda_W^2)4\sum_iC_id_i
[L_1(p^2,m_{dbi1},m_{dbi2})}\nonumber\\
&&+L_2(p^2,m_{dbi1},m_{dbi2})
+L_2(p^2,m_{dbi2},m_{dbi1})]\}^{1/2}.\nonumber\\\label{Lambda}
\end{eqnarray}
The $m_{dbi}$ denote the fermion doublets: $[e,\nu_e],
[\mu,\nu_\mu], [\tau, \nu_\tau], [d,u], [s,c], [b,t]$, and the
color factors are $C_\ell=1$ for leptons and $C_q=3$ for quarks.

The quantities that enter the calculations of $\rho$ and
$\Lambda_W$ are obtained from the traces of (\ref{W3}) and
(\ref{Wcharged}). We have
\begin{equation}
L_1=\frac{1}{4}(3L_{1+}+L_{1-}),\quad
L_2=\frac{1}{4}(3L_{2+}+L_{2-}).
\end{equation}

The $Ls$ are given by
\begin{eqnarray}
&&L_{1\pm}(p^2,m_1,m_2)\nonumber\\
&&=\int^\infty_1 d\tau_1\int^\infty_1
d\tau_2\exp\biggl[-\frac{\tau_1m_1^2+\tau_2m_2^2}{\Lambda_W^2}
-\frac{\tau_1\tau_2
p^2}{(\tau_1+\tau_2)\Lambda_W^2}\biggr]\nonumber\\
&&\times\biggl(\frac{\Lambda_W^2}{(\tau_1+\tau_2)^3}
\pm\frac{p^2\tau_1\tau_2}{(\tau_1+\tau_2)^4}\biggr),
\end{eqnarray}
\begin{eqnarray}
&&L_{2\pm}(p^2,m_1,m_2)\nonumber\\
&&=\int^1_0 d\tau_1\int^\infty_1
d\tau_2\exp\biggl[-\frac{\tau_1m_1^2+\tau_2m_2^2}{\Lambda_W^2}
-\frac{\tau_1\tau_2
p^2}{(\tau_1+\tau_2)\Lambda_W^2}\biggr]\nonumber\\
&&\times\biggl(\frac{\Lambda_W^2}{(\tau_1+\tau_2)^3}
\pm\frac{p^2\tau_1\tau_2}{(\tau_1+\tau_2)^4}\biggr).
\end{eqnarray}

In the calculation of $\Lambda_W$ and $\rho$, we include the six
observed lepton masses~\cite{pdg}:
\begin{eqnarray}
&&m_e=0.000511\, {\rm GeV},\, m_\mu=0.10566\, {\rm GeV},\nonumber\\
&&m_\tau=1.777\pm 0.025\, {\rm GeV},\, m_{\nu_e}=0.2\times
10^{-8}\pm 0.09\, {\rm GeV},\nonumber\\
&&m_{\nu_\mu}=0.00019\pm 0.07\, {\rm GeV},\, m_{\nu_\tau}=0.0182\pm
1.8\, {\rm GeV},\nonumber\\
\end{eqnarray}
and the six observed quark masses~\cite{pdg,Tevatron}:
\begin{eqnarray}
&&m_u=0.0019\, {\rm GeV},\, m_d=0.0044\, {\rm GeV},\nonumber\\
&&m_s=0.095\pm 0.025\, {\rm GeV},\, m_c=1.31\pm 0.09\, {\rm
GeV},\nonumber\\
&&m_b=4.24\pm 0.07\, {\rm GeV},\, m_t=170.9\pm 1.8\, {\rm GeV}.
\end{eqnarray}

From (\ref{rhoeq}), it follows that for $m_f=0$ we get $\rho=1$
and for $m_t\rightarrow \infty$ we obtain $\rho=0.857$. The former
result corresponds to the tree graph value obtained from the
standard model and yields the prediction:
\begin{equation}
s_Z^2\equiv \sin^2\theta_W(M_Z)=0.22239\pm 0.00095.
\end{equation}
In the calculations of $\rho$ and $\Lambda_W$ the top quark mass
$m_t$ dominates the calculations.

We obtain from (\ref{Lambda}) the non-local energy scale at the
Euclidean on shell value $p^2=M_Z^2$:
\begin{equation}
\label{LambdaW} \Lambda_W(M_Z)=541\, {\rm GeV},
\end{equation}
and for $\rho$ at the Z-pole, we get from (\ref{rhoeq}):
\begin{equation}
\label{rhovalue} \rho(M_Z)=0.993.
\end{equation}
The experimental values of the $W$ and $Z$ boson masses are given
by~\cite{Tevatron,pdg}:
\begin{equation}
M_W=80.413\pm 0.029\, {\rm GeV},\, M_Z=91.1876\pm 0.0021\, {\rm
GeV}.
\end{equation}
This predicts from (\ref{rho}) and (\ref{rhovalue}) the value
\begin{equation}
\label{sin2} s_Z^2\equiv \sin^2\theta_W(M_Z)=0.21686\pm 0.00097.
\end{equation}

A calculation in the standard model including radiative
corrections gives the value~\cite{pdg}:
\begin{equation}
s_Z^2=0.23122\pm 0.00015.
\end{equation}
A more accurate prediction of $s_Z^2$ in our non-local QFT
requires calculating further radiative corrections in our
electroweak model. A calculation of the $\rho$ and $\Lambda_W$ at
$p^2=0$ yields values close to (\ref{LambdaW}) and (\ref{sin2})
calculated at the $Z$-pole.

\section{Fermion Masses}

In the standard electroweak model, fermion masses are generated
through Yukawa couplings $\langle\phi\rangle\bar\psi\psi$ where
$\langle\phi\rangle$ is the vacuum expectation value of the Higgs
field $\phi$. In our model, we shall generate fermion masses from
the finite one-loop fermion self-energy graph by means of a
Nambu-Jona-Lasinio mechanism~\cite{Nambu}. The one-loop fermion
self-energy graphs are shown in Figure 4. A fermion particle will
satisfy
\begin{equation}
\label{meq} i\gamma\cdot p+m_{0f}+\Sigma(p)=0,
\end{equation}
for $i\gamma\cdot p+m_f=0$ where $m_{0f}$ is the bare fermion
mass, $m_f$ is the observed fermion mass and $\Sigma(p)$ is the
{\it finite} proper self-energy part. We have
\begin{equation}
\label{sigmaeq}
m_f-m_{0f}=\Sigma(p,m_f,g,\Lambda_f)\vert_{i\gamma\cdot p+m_f=0}.
\end{equation}
Here, $\Lambda_f$ denote the non-local energy scales for lepton
and quark masses. We can solve (\ref{meq}) and (\ref{sigmaeq}) by
successive approximations starting from the bare mass $m_{0f}$.
However, we can also find solutions for $m_f\not= 0$ when
$m_{0f}=0$ for a broken symmetry vacuum state.

The finite self-energy contribution obtained by joining together
two fermion-boson vertices is given by~\cite{Moffat5}:
\begin{eqnarray}
\lefteqn{-i\Sigma_1(p)=\int
\frac{d^4k}{(2\pi)^2}(ig^2_f\gamma^\mu)\biggl(\frac{-i}{\gamma\cdot\partial
+m_f-i\epsilon}\biggr)(ig^2_f\gamma^\nu)}\nonumber\\
&&\times\biggl(\frac{-i\eta_{\mu\nu}}{k^2+m^2-i\epsilon}\biggr)\nonumber\\
&&\times\exp\biggl[-\biggl(\frac{p^2+m_f^2}{\Lambda_f^2}\biggr)
-\biggl(\frac{q^2+m_f^2}{\Lambda_f^2}\biggr)-\frac{k^2}{\Lambda_f^2}\biggr],
\end{eqnarray}
where $g^2_f$ is a fermion coupling constant which contains quark
color factors, $q=p-k$ and we choose the boson zero mass limit.
The propagators are now converted to Schwinger integrals and the
momentum integration is performed to give
\begin{eqnarray}
\lefteqn{-i\Sigma_1(p)
=-g^2_f\exp\biggl[-\biggl(\frac{p^2+m_f^2}{\Lambda_f^2}\biggr)\biggr]}\nonumber\\
&&\times\int^\infty_1\frac{d\tau_1}{\Lambda_f^2}\int^\infty_1\frac{d\tau_2}{\Lambda_f^2}\int\frac{d^4k}{(2\pi)^4}
(2\gamma\cdot q+4m_f)\nonumber\\
&&\times\exp\biggl[-\tau_1\biggl(\frac{q^2+m^2}{\Lambda_f^2}\biggr)-\tau_2\frac{k^2}{\Lambda_f^2}\biggr]\nonumber\\
&&=\frac{-ig_f^2}{8\pi^2}\exp\biggl[-\biggl(\frac{p^2+m_f^2}{\Lambda_f^2}\biggr)\biggr]\nonumber\\
&&\times\int^\infty_1 d\tau_1\int^\infty_1
d\tau_2\biggl[\frac{\tau_2}{(\tau_1+\tau_2)^3}\gamma\cdot
p+\frac{2m_f}{(\tau_1+\tau_2)^2}\biggr]\nonumber\\
&&\times\exp\biggl(-\frac{\tau_1\tau_2}{\tau_1+\tau_2}\frac{p^2}{\Lambda_f^2}
-\tau_1\frac{m_f^2}{\Lambda_f^2}\biggr).
\end{eqnarray}
Here, we have performed a rotation to Euclidean momentum space,
accounting for the factor of $i$.

Another contribution $\Sigma_2(p)$ to the self-energy comes from
the tadpole fermion-boson self-energy graph:
\begin{eqnarray}
\lefteqn{-i\Sigma_2(p)=\int\frac{d^4k}{(2\pi)^2}(-ig_f^2)\gamma^\mu(\gamma\cdot
q-m_f)\gamma^\nu\biggl(\frac{-i\eta_{\mu\nu}}{k^2-i\epsilon}\biggr)}\nonumber\\
&&\times\int_0^1\frac{d\tau}{\Lambda_f^2}\exp\biggl[-\biggl(\frac{p^2+m_f^2}{\Lambda_f^2}\biggr)
-\tau\biggl(\frac{q^2+m_f^2}{\Lambda_f^2}\biggr)-\frac{k^2}{\Lambda_f^2}\biggr].\nonumber\\
\end{eqnarray}

Adding together the two diagram contributions $\Sigma_1(p)$ and
$\Sigma_2(p)$, we obtain
\begin{eqnarray}
\lefteqn{\Sigma(p)=\frac{g^2_f}{8\pi^2}\exp\biggl[-\biggl(\frac{p^2+m_f^2}{\Lambda_f^2}\biggr)\biggr]}\nonumber\\
&&\times\int^1_0 dx(x\gamma\cdot
p+2m_f)E_1\biggl[(1-x)\frac{p^2}{\Lambda^2_f}+\biggl(\frac{1-x}{x}\biggr)\frac{m_f^2}{\Lambda_f^2}\biggr].\nonumber\\
\end{eqnarray}
Here, $E_1$ is the exponential integral:
\begin{eqnarray}
\lefteqn{E_1(z)\equiv \int^\infty_z dy\frac{\exp(-y)}{y}}\nonumber\\
&&=-\ln(z)-\gamma_e-\sum^\infty_{n=1}\frac{(-z)^n}{nn!},
\end{eqnarray}
where $\gamma_e$ is Euler's constant. By developing an asymptotic
expansion in $\Lambda_f$ and expanding the exponential integral,
we get
\begin{eqnarray}
\lefteqn{\Sigma(p)=\frac{\alpha_f}{2\pi}\biggl[\biggl(\frac{1}{2}\gamma\cdot\partial
p+2m_f\biggr)\ln(\Lambda_f^2)}\nonumber\\
&&-\biggl(\frac{1}{2}\gamma\cdot\partial p+2m_f\biggr)\gamma_e
+\frac{1}{2}\gamma\cdot\partial p\nonumber\\
&&-\int^1_0 dx(x\gamma\cdot\partial
p+2m_f)\ln(xp^2+m_f^2)\nonumber\\
&&+O\biggl[\frac{\ln(\Lambda^2_f)}{\Lambda_f^2}\biggr],
\end{eqnarray}
where $\alpha_f=g^2_f/4\pi$.

\begin{figure}
\includegraphics[width=1.0\linewidth]{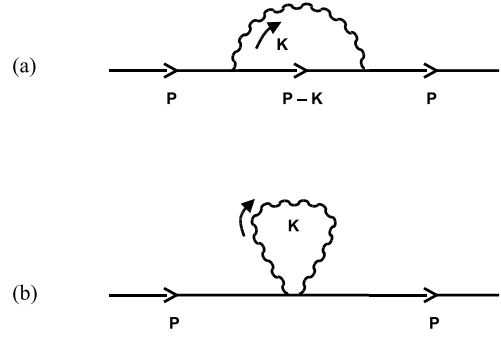}
\caption{(a) the one-loop fermion  self-energy graph; (b) the
fermion one-loop self-energy tadpole graph.} \label{fig:4}
\end{figure}

The fermion mass is now identified with $\Sigma(p)$ at $p=0$:
\begin{equation}
m_f=\Sigma(0)=\frac{\alpha_fm_f}{\pi}\biggl[\ln\biggl(\frac{\Lambda_f^2}{m_f^2}\biggr)
-\gamma_e\biggr]
+O\biggl[\frac{\ln(\Lambda^2_f)}{\Lambda_f^2}\biggr].
\end{equation}
This equation has two solutions: either $m_f=0$, or
\begin{equation}
1=\frac{\alpha_f}{\pi}\biggl[\ln\biggl(\frac{\Lambda_f^2}{m_f^2}\biggr)-\gamma_e\biggr].
\end{equation}
The first trivial solution corresponds to the standard
perturbation result. The second non-trivial solution will
determine $m_f$ in terms of $\alpha_f$ and $\Lambda_f$ and leads
to the fermion ``mass gap'' equation
\begin{equation}
m_f=\Lambda_f\exp\biggl[-\frac{1}{2}\biggl(\frac{\pi}{\alpha_f}+\gamma_e\biggr)\biggr].
\end{equation}
The non-perturbative solution for the fermion masses is based on a
broken vacuum state with $\langle{\bar\psi\psi}\rangle_0\not=0$
and avoids introducing bare fermion masses in the $SU(2)\times
U(1)$ gauge invariant zero mass and zero interaction limit. The
Lagrangian picks up fermion mass terms:
\begin{eqnarray}
\lefteqn{L_{M_f}=m_f{\bar\psi}_f\psi_f}\nonumber\\
&=&m_f\psi_f[\frac{1}{2}(1-\gamma_5)+\frac{1}{2}(1+\gamma_5)]\psi_f\nonumber\\
&=&m_f({\bar\psi}^R\psi^L+{\bar\psi}^L\psi^R).
\end{eqnarray}

By choosing $\alpha_f =\alpha_s(M_Z)=0.117\pm 0.0007$, where
$\alpha_s$ is the strong coupling constant evaluated at the $m_Z$
pole, and $\Lambda_t=1.55\times 10^5\,{\rm TeV}$, we find for the
top quark mass, $m_t=171\, {\rm GeV}$.

\section{Unitarity and the Scale of $W_LW_L$ Scattering}

We can separate the Lagrangian into a gauge invariant sector and a
symmetry breaking sector
\begin{equation}
{\cal L}={\cal L}_{\rm gauge\, inv}+{\cal L}_{\rm SB}.
\end{equation}
Here, ${\cal L}_{\rm gauge\, inv}$ possesses the unbroken
$G(NL,4)$ gauge symmetry with massless, transversely polarized
gauge bosons $W, Z$ and $\gamma$. ${\cal L}_{\rm SB}$ describes
the dynamical symmetry breaking. In the standard model, the Higgs
mechanism ensures that the three Goldstone bosons, $w^\pm$ and
$z$, couple to the three gauge currents associated with the
spontaneously broken symmetry of $SU(2)_L\times U(1)_Y$. At the
same time, it ensures that $w^\pm$ and $z$ become longitudinal
modes of the gauge bosons $W^\pm$ and $Z$, which become massive
independently of whether the Higgs boson exists or not.

A rigorous bound on the energy dependence of $W_LW_L$ scattering
comes from unitarity. If we set $\rho=1$, then the $I=J=0$ partial
wave amplitude is given by
\begin{equation}
A_{00}(W_LW_L)=\frac{s}{16\pi v^2},
\end{equation}
where $s$ is the square of the center-of-mass energy. Unitarity
demands that below 4-body thresholds $A_{00}\leq 1$, and
$Re(A_{00}) \leq 1/2$. Both of these conditions are violated above
1.8 and 1.2 TeV, respectively~\cite{Chanowitz}.

For $s\ << M^2_{\rm SB}$, where $M_{\rm SB}$ is the typical
symmetry breaking mass scale of ${\cal L}_{\rm SB}$, we get the
amplitude:
\begin{equation}
{\cal A}(W^+_LW^-_L\rightarrow
Z_LZ_L)=\frac{1}{\rho}\frac{s}{v^2},
\end{equation}
in the energy range $M_W^2 \ll s \ll M^2_{\rm SB}$. For leading
order in the unitary-gauge the amplitude in the gauge sector
yields the behavior:
\begin{equation}
{\cal A}(W^+_LW^-_L\rightarrow Z_LZ_L)_{\rm gauge\, sector}
$$ $$
=\frac{g^2s}{4\rho M_W^2}+O(g^2s^0)\sim \frac{s}{\rho v^2}.
\end{equation}
This high energy behavior eventually violates unitarity and the
standard local QFT theory without a Higgs particle is
non-renormalizable. The inclusion of a Higgs exchange at the scale
$M_{\rm SB}$ cancels the ``bad'' high energy behavior and allows
unitarity to be satisfied above $\sim\, 1$ TeV and the theory is
renormalizable. In particular, the low energy behavior for $s\ll
M^2_{\rm SB}$ can be shown to decouple the ${\cal L}_{\rm SB}$ to
all orders.

For our finite non-local electroweak theory, we require that
${\cal A}(W^+_LW^-_L\rightarrow Z_LZ_L)$ {\it vanishes} above
$\sim\, 1$ TeV, avoiding the unitarity violating behavior of the
standard gauge sector with massive vector bosons $W$ and $Z$. We
implement this by requiring that the symmetry breaking fermion
loop measure becomes the gauge invariant measure, $\mu_{\rm
SB}\rightarrow \mu_{\rm inv}$, above $\sim\, 1$ TeV. Thus, the $W$
and $Z$ bosons become massless gauge bosons above $\sim\, 1$ TeV
with only transverse polarization degrees of freedom and there is
no violation of unitarity. The massless boson, non-local gauge invariant
theory becomes the standard local renormalizable theory when
$\Lambda_W\rightarrow \infty$. The signature of a vanishing cross
section for $W_L^+W_L^-$ scattering above $\sim\, 1$ TeV should be
detectable at the LHC.

In the non-local QFT with a finite non-local scale $\Lambda_W$,
the partial wave scattering amplitude $A_\ell(s,t)$ (where $t$ is
the momentum transfer squared) {\it for the crossed channel} in
lowest order scattering diagrams will behave badly at high
energies. A similar phenomenon occurs in perturbative string
theory~\cite{Gross}. To circumvent this problem, it is necessary
to re-sum the scattering amplitudes to arbitrary order in
perturbation theory. In QFT the on-shell high energy behavior of
scattering amplitudes poses difficult questions, for it combines
both short and long distance physics. The fixed $t$, large $s$
behavior of QCD reveals this problem even for the case of fixed
angle scattering. For FQFT the {\it exact} leading behavior of the
scattering amplitude has to be deduced, order by order in
perturbation theory, by means of a saddle point calculation. The
exponential behavior of string theory scattering amplitudes is
unlike the power behavior that holds in QFT. The same is true of
the finite loop graphs in non-local QFT. In contrast to
perturbative string theory, the FQFT tree graphs behave at high
energies as in local, point particle QFT. However, the lowest
order behavior of the crossed channel amplitudes in non-local QFT
violates the rigorous bound of Cerulus and Martin~\cite{Martin},
which states that $\vert A(s,\cos\theta)\vert
\geq\exp[-\sqrt{s}\ln s\, c(\theta)]$. The proof of this bound
uses unitarity, a finite mass threshold gap and the assumption of
a polynomial bound in the energy for fixed $t$ of the scattering
amplitudes. The polynomial behavior of scattering amplitudes in
standard QFT is a consequence of {\it locality}. The non-local QFT
like string theory manages to be sufficiently non-local to avoid a
polynomial bound, yet maintains sufficiently local interactions to
preserve causality.

\section{Experimental Signatures of Non-local Electroweak Theory}

We have seen that at $\sqrt{s}\sim\, 0.5-1$ TeV the non-local
scale $\Lambda_W$ becomes significant corresponding to an
exponential fall-off of the loop scattering amplitudes in the
s-channel. The tree graph scattering amplitudes at large energies
are strictly local and have the same high energy behavior as in
the standard point QFT. A detection of a non-local behavior in the
loop scattering amplitudes at the LHC at $\sqrt{s} > 0.5-1$ TeV
would be an experimental confirmation of non-local QFT. This could
be observed in a different high energy behavior of scattering
amplitudes involving loop diagrams in, say, photon-proton Compton
scattering.

In the standard electroweak model, the quadratic divergence of the
Higgs mass radiative correction is a serious problem and has
motivated several alternative models beyond the standard model. In
the Higgs potential:
\begin{equation}
V(\phi)=-\frac{1}{2}\mu^2\phi^2+\frac{1}{4}\lambda\phi^4,
\end{equation}
the mass parameter $\mu^2$ at tree level ($\mu^2 > 0$ for
spontaneous symmetry breaking) is related to the vacuum
expectation value $v$ by $\mu^2=\lambda v^2$. When radiative
corrections are calculated, the mass parameter becomes
$\mu_0^2+\delta\mu^2$, where $\mu_0^2$ is the bare mass term and
$\delta\mu^2$ is the radiative correction. Since the radiative
correction depends on the cutoff scale as $\Lambda_C^2$, the fine-tuning
between the bare mass term and the radiative correction
becomes significant for $\Lambda_C > 1$ TeV. If the cutoff reaches
the Planck energy scale, then the amount of fine-tuning is
enormous and a tuning of order $10^{32}$ is needed. This is the
source of the naturalness or hierachy problem in standard
electroweak theory. Because our FQFT does not possess a Higgs
particle, the Higgs fine-tuning problem does not occur.

Because the $W$, $Z$ and $\gamma$ tree graphs in FQFT electroweak
theory are identical to those of the standard electroweak theory,
all the lower energy predictions of the standard model at tree
graph level remain the same when the Higgs particle graphs are
excluded. One probe of the FQFT is the calculated difference
between the experimental and local, standard model theoretical
values of the muon $a_\mu=(g-2)/2$ anomaly. The Brookhaven
National Laboratory experiment has determined $a_\mu$ with a much
improved accuracy~\cite{BNL}. The experimental value is:
$(a_\mu)_{\rm exp}=1165208.0(6.3)\times 10^{-10}$. The difference
between this result and the standard model prediction $(a_\mu)_\mathrm{SM}$ is:
$(22.4\pm 10)\times 10^{-10}\le \delta a_\mu \le (26.1\pm 9.4)\times 10^{-10}$.

We obtain the bound on a possible non-local QFT contribution for
the anomalous magnetic moment of the muon:
\begin{equation}
\delta(a_\mu)_{\rm nl}\le\frac{K}{\Lambda_W^2}.
\label{expbound}
\end{equation}
From (\ref{expbound}) and from our estimated value for the
electroweak non-local scale $\Lambda_W(M_Z)=541$ GeV, we get
the bound:
\begin{equation}
K \leq 701\pm 338\,({\rm MeV})^2.
\end{equation}
Thus, a careful calculation of the non-local contribution to
$a_\mu$ could test the prediction of FQFT.

\section{Conclusions}

We have constructed an electroweak model based on a method to make
a massless gauge invariant QFT into a non-local theory which is
finite, Poincar\'e invariant, and perturbatively unitary. The
method has two stages -- classical and quantum. In the first we
make the theory finite by non-localizing its interactions. The
violation of unphysical decoupling is then removed at each order
by adding an appropriate new interaction. The resulting tree
graphs decouple from unphysical modes and the action possesses a
generalized gauge invariance in the form of the non-local group
$G(NL,4)$. In the electroweak theory, the new symmetry can be
viewed on shell as a non-local and non-linear representation of
$SU(2)\times U(1)$.

The quantum stage of our method consists of finding measures
to make the functional formalism invariant, and then to
find a path integral measure that dynamically breaks the non-local gauge
symmetry to give the $W$ and $Z$ bosons mass while keeping the
photon massless. We have also proposed a method for giving leptons
and quarks mass through a mass gap equation. This is done directly
through the finite fermion self-energy radiative diagram in terms
of a fermion mass scale $\Lambda_f$.

The number of unknown parameters in our extended electroweak
theory is reduced by not having an unpredictable Higgs mass but we
have a new predicted energy scale parameter $\Lambda_W$. The
number of fermion mass scale parameters $\Lambda_f$ -- one for each
observed lepton and quark -- is the same as occurs in the standard
Higgs Yukawa coupling model. This number of undetermined
parameters still points to the need for a more comprehensive
unified theory of the particle interactions, which would determine
the unknown parameters in a fundamental way.

The nice feature of our extended electroweak theory is that it
does not increase the number of particles, nor does it extend the
number of dimensions as in string theory, yet it preserves
Poincar\'e invariance and a finite electroweak theory.

We have proposed ways to test our FQFT. The vanishing of the cross
section for $W^+_LW^-_L\rightarrow W^+_LW^-_L$ above $\sim\, 1$
TeV should be observable at the LHC. Moreover, the
behavior of scattering amplitudes for, say, Compton proton-photon
scattering above $\Lambda_W\sim 0.5-1$ TeV is another way to
detect the non-local behavior of finite loop graphs. We have also
shown that a calculation of the non-local loop contribution to the
muon anomalous $g-2$ magnetic moment $a_\mu$ could reveal the
difference between FQFT with $\Lambda_W\sim 0.5$ TeV and the
standard model calculation of $a_\mu$.

If the Tevatron and LHC accelerator experiments fail to detect a
Higgs particle, then a physically consistent theory of electroweak
symmetry breaking such as the one studied here, in which no Higgs
particle is included in the particle spectrum, will be required to
understand the important phenomenon of how the $W$ and $Z$ bosons
and fermions acquire mass. \vskip 0.2 true in

{\bf Acknowledgements:}  I thank Clifford Burgess and Martin Green
for stimulating discussions. I also thank Viktor Toth for
stimulating discussions and valuable suggestions for obtaining and
checking numerical results. This research was supported by a grant
from NSERC. The Perimeter Institute is supported in part by the
Government of Canada through NSERC and by the Province of Ontario
through MEDT.

\label{lastpage}


\begin{thebibliography}{99}

\bibitem{Higgs} P. Higgs, Phys. Lett. {\bf 12}, 132 (1964); P. Higgs, Phys. Rev. Lett. {\bf 13}, 508 (1964);
F. Englert and R. Brout, Phys. Rev. Lett. {\bf 13}, 321 (1964); P.
Higgs, Phys. Rev. {\bf 145}, 1156 (1966).

\bibitem{Halzen} F. Halzen and A. D. Martin, Quarks and Leptons:
An Introductory Course in Modern Particle Physics, John Wiley \&
Sons. 1984.

\bibitem{Burgess} C. P. Burgess and G. D. Moore, The Standard Model: A Primer, Cambridge
University Press, 2007.

\bibitem{CERN} The LEP Electroweak Working Group,
http://lepewwg.web.cern.ch

\bibitem{Moffat} J. W. Moffat, Phys. Rev. D{\bf 41}, 1177
(1990).

\bibitem{Moffat2} J. W. Moffat, Mod. Phys. Lett. {\bf 6}, 1011
(1991).

\bibitem{Moffat3} M. Clayton and J. W. Moffat, Mod. Phys. Lett. {\bf 6},
2697 (1991).

\bibitem{Moffat4} J. W. Moffat, Phys. Rev. D{\bf 39}, 3654 (1989).

\bibitem{Moffat5} D. Evens, J. W. Moffat, G. Kleppe and R. P. Woodard,
Phys. Rev. D{\bf43}, 499 (1991).

\bibitem{Moffat6} J. W. Moffat and S. M. Robbins, Mod. Phys. Lett.
A{\bf 6}, 1581 (1991).

\bibitem{Hand} B. J. Hand and J. W. Moffat, Phys. Rev. {\bf D43},
1896 (1991).

\bibitem{Woodard} G. Kleppe and R. P. Woodard, Phys. Lett. B{\bf 253},
331 (1991).

\bibitem{Kleppe} G. Kleppe and R. P. Woodard, Nucl. Phys. B{\bf 388},
81 (1992).

\bibitem{Hand2} B. J. Hand, Phys. Lett. {\bf B275}, 419 (1992).

\bibitem{Cornish} N. J. Cornish, Mod. Phys.
Lett. {\bf 7}, 631 (1992).

\bibitem{Cornish2} N. J. Cornish, Mod. Phys. Lett. {\bf 7}, 1895
(1992).

\bibitem{Cornish3} N. J. Cornish, Int. J. Mod. Phys. A {\bf 7},
6121 (1992).

\bibitem{Woodard2} G. Kleppe and R. P. Woodard, Ann. of  Phys.
{\bf 221}, 106 (1993).

\bibitem{Clayton} M. A. Clayton, L. Demopolous and J. W. Moffat,
Int. J. Mod. Phys. A{\bf 9}, 4549 (1994).

\bibitem{Paris} J. Paris, Nucl. Phys. {\bf B450}, 357 (1995).

\bibitem{Troost} J. Paris and W. Troost, Nucl. Phys. {\bf B482}.
373 (1996).

\bibitem{Joglekar} G. Saini and S. D. Joglekar, Z. Phys. {\bf
C76}, 343 (1997).

\bibitem{Moffat7} J. W. Moffat, arXiv: hep-ph/0003171 v2; arXiv:
hep-ph/0102088 v2.

\bibitem{Tevatron} Tevatron Electroweak Working Group
(http://tevewwg.fnal.gov/top/).

\bibitem{pdg} W.-M. Yao, et al., Review of Particle Physics, J. of
Phys., (http://pdg.lbl.gov.)

\bibitem{Nambu} Y. Nambu and G. Jona-Lasinio, Phys. Rev. {\bf
122}, 345 (1961).

\bibitem{Chanowitz} M. S. Chanowitz, Czech. J. Phys. {\bf 55}, B45
(2005), ArXiv: hep-ph/0412203.

\bibitem{Gross} D. J. Gross and P. F. Mende, Phys. Lett. {\bf
B197}, 129 (1987).

\bibitem{Martin} F. Cerulus and A. Martin, Phys. Lett. {\bf 8}, 80
(1963).

\bibitem{BNL} G. W. Bennett et al., Phys. Rev. {\bf D73} 072003
(2006).

\end{thebibliography}
\end{document}